# The structural constraints of income inequality in Latin America

Dominik Hartmann[1,2], Cristian Jara-Figueroa[1], Miguel Guevara[3], Alex Simoes[4] and César A. Hidalgo[1]

AFFILIATIONS:
[1] Macro Connections, The MIT Media Lab, Massachusetts Institute of Technology, US
[2] Chair for Economics of Innovation, University of Hohenheim, DE.
[3] Universidad de Playa Ancha, Valparaíso, CL
[4] Datawheel, Cambridge-MA, US

**Abstract**: Recent work has shown that a country's productive structure constrains its level of economic growth and income inequality. Here, we compare the productive structure of countries in Latin America and the Caribbean (LAC) with that of China and other High-Performing Asian Economies (HPAE) to expose the increasing gap in their productive capabilities. Moreover, we use the product space and the Product Gini Index to reveal the structural constraints on income inequality. Our network maps reveal that HPAE have managed to diversify into products typically produced by countries with low levels of income inequality, while LAC economies have remained dependent on products related to high levels of income inequality. We also introduce the *Xgini*, a coefficient that captures the constraints on income inequality imposed by the mix of products a country makes. Finally, we argue that LAC countries need to emphasize a smart combination of social and economic policies to overcome the structural constraints for inclusive growth.

## 1 Introduction

Decades ago, in the 1950s and 1960s, Development Pioneers and Latin American Structuralists argued that the productive structure of a country constrains its ability to generate and distribute income (Rosenstein-Rodan 1943; Prebisch 1949; Singer 1950; Hirschman 1958; Furtado 1959). While the focus on productive structures waned in the 1980s and 1990s, it was recently revived by research that showed how the mix of products that a country exports is predictive of its future pattern of diversification (Hidalgo et al. 2007), economic growth (Hidalgo and Hausmann 2009; Hausmann et al. 2014), and income inequality (Hartmann et al. 2015). This new line of research on economic complexity focuses on the ability of economies to produce a diverse and sophisticated mix of products.

In this paper, we compare the productive sophistication and structural constraints on income inequality of countries in Latin American and the Caribbean (LAC) with that of China and other high-performing Asian Economies (HPAE), such as South Korea, Singapore, or Malaysia. The results show a large gap in the productive capabilities of LAC and HPAE that has been significantly increasing since 1990. Moreover, we use the product space and the Product Gini Index (Hartmann et al. 2015) to reveal how changes in the productive structure translate into changes in opportunities for inequality reduction. The network illustrates how HPAE have managed to diversify into more sophisticated industrial products which are typically produced by countries with low levels of income inequality, such as electronics and machinery. Conversely, the productive portfolio of LAC countries has remained largely dependent on products related with high levels of income inequality, such as crude petroleum, copper, and coffee beans. We also introduce the *Xgini*, a coefficient which estimates the expected level of inequality associated to type of products a country makes and is able capture the constraints on income inequality imposed by a country's productive structure. While the Xginis of LAC countries have remained at a very high level, indicating strong constraints on inequality reduction, the Xginis of many HPAEs have declined significantly. This means while HPAE have opened up many opportunities for inclusive growth and reduction of income inequality, LAC's productive structure strongly constrains it ability to generate and distribute income. From an economic policy perspective, we argue that LAC countries need to emphasize and move towards a smart combination of social and economic policies, rather than continuing the state versus market debate, in order to overcome their structural constraints on inclusive growth.

**The Latin America development policy debate**

Debates connecting productive structures, economic growth, and income inequality have a long academic tradition, especially in Latin America. During the second half of the twentieth century Latin America was at the center of the discussion about development theories favoring free markets (Kuznets 1955; Krueger 1985) and theories promoting state intervention (Rosenstein-Rodan 1943; Prebisch 1949; Furtado 1959; Sunkel and Girvan 1973).

The state versus market debate had profound consequences in Latin America, as economic policies based on both sides of the debate were implemented at different times. The industrialization efforts in Brazil, during the 1960s, 1970s, and 1980s, or the efforts to deregulate and liberalize the economy in Chile during the 1980s are clear examples of both strategies. However, both approaches had serious shortcomings. On the one hand, "state driven industrialization through import-substitution" led to major economic inefficiencies and the so-called "lost decade". On the other hand. the wave of market liberalization and structural reforms, adopted in the 1990s, stabilized the economy and generated economic growth, but failed to create an inclusive economy needed to integrate Latin America's poorest citizens.

The failure of both extreme positions has led economists to consider a middle ground favoring a congruent mix of social and economic policies to promote innovation and economic complexity (Wade 1990; Rodrik 2004; Hartmann 2014). This middle ground makes use of the strengths of both approaches by overcoming both market and government failures, investing in human capital, and promoting economic self-discovery processes and international innovation linkages (Rodrik 2004; Hartmann 2014). Several HPAEs experienced rapid economic development under policies that mixed markets and targeted state intervention (Wade 1990; Stiglitz 1996; Rodrik 2004; Hartmann 2014). Their economic success has been a motivation for a middle ground, mixing social and economic policies. Conversely, this strategic middle ground has not yet still to be implemented in Latin America, which is to a great extent still divided between factions that still believe that excessive state intervention is the solution to the regions problems, and factions that believe that complete liberalization and deregulation is the only way for economies to move forward.

In the meantime, the commodity boom, and the rise in the price of natural resources have provided LAC countries with the economic resources needed to implement social policy programs—like conditional cash-transfer programs or higher expenditures in education and health. These programs have led to a significant reduction in poverty and an increase in the region's level of human development (Hartmann 2011). Nonetheless the commodity boom did not translate into a substantial increase in the capacity of LAC economies to produce more sophisticated products. In fact during the commodity and natural resources prices boom, many Latin American economies fell behind in terms of economic complexity. According to MIT's Economic Complexity Ranking, Brazil fell from position 29 to 56 between 1990 and 2013[1]. Chile, which has been the economic darling of the region due to its economic growth, went from 54 to 67 between 1990 and 2013. As commodity prices decrease and the resources needed to support social programs become scarcer, Latin America once again finds itself in a predicament.

**Economic complexity, institutions, and income inequality**

In this paper, we compare the evolution of economic complexity and the related structural constraints on the reduction of income inequality in LAC and HPAE. The connection between the complexity of a country's economy and its level of income inequality can be understood by interpreting the industries present in the country as the embodiment of the many factors that make economies prosperous and inclusive. The industries present in an economy tell us about the knowledge embodied in its population, the job opportunities and bargaining power of workers, the industrial sectors that the economy can diversify into, and the quality of its institutions (Hartmann et al. 2015). For example, complex industries, such as advanced medical equipment or software development, require better-educated and more creative workers, and moreover institutions that are able to include the creative inputs of workers into the activities of firms. In consequence, an economy's productive matrix can be seen as a proxy for a number of explanatory factors, such as the productive knowledge and the inclusiveness of institutions, that profoundly affect economic growth and inequality, but that are typically difficult to measure directly.

---

[1] http://atlas.media.mit.edu/en/rankings/country/1990/
and http://atlas.media.mit.edu/en/rankings/country/2013/

The close relationship between an economy's industries and its institutions implies that social policies alone might lack the strength required to modify a country's level of income inequality beyond the range that is typically expected given its productive structure. Therefore, industrial policies need to compliment social policies in order to achieve a substantial change (Amsden 2010; Hartmann 2014).

In this article, we use methods from network science, economic complexity research, and data visualization to show LAC's structural constraints on economic growth and inequality reduction. These methods enable a more detailed picture, or fingerprint, of the economy, revealing the knowledge landscape and economic opportunities of countries and regions. Moreover, these methods allow for a more detailed comparison between LAC economies with high performing Asian economies (HPAE).

The remainder of the paper is structured as follows. Section 2 introduces the Data and Methods. Section 3 compares the economic complexity and structural transformation of LAC and HPAE, showing a large gap in productive capabilities and know-how of both regions. Section 4 then illustrates how the productive structures of LAC constrain their possibilities for inclusive growth and inequality reduction. Section 5 interprets the empirical results from an economic policy perspective, and highlights the need for establishing prolific innovation systems to overcome LAC's structural economic constraints. Section 6 provides concluding remarks.

## 2   Data and Methods

We use data on world trade, economic complexity, and income inequality to compare the structural constraints of LAC and HPAE. Data on income inequality comes from the Galbraith et al., 2014 (GINI EHII dataset). Due to the sparseness of the Gini data, we interpolate the missing years using linear splines. Moreover, we consider only the countries for which the Economic Complexity Index is available. The data on world trade, compiled by Feenstra et al. (2005), combines exports data from 1962 to 2000 with data from the U.N. Comtrade from the period between 2001 and 2012. The values for the Economic Complexity Index come from MIT's Observatory of Economic Complexity (atlas.media.mit.edu) (Simoes and Hidalgo 2011).

We use the Economic Complexity Index (ECI) as indicator for the know-how and productive capabilities of LAC and HPAE countries. ECI measures the sophistication of a country's productive structure, combining information on the diversity and ubiquity of the products a country's exports (Hidalgo and Hausmann 2009). The intuition behind ECI is that sophisticated economies are diverse and export products produced by few other economies. ECI can be interpreted as a measure of a country's productive capabilities that are embodied in its institutions and people. Further information about the calculation of ECI can be found in Hidalgo and Hausmann (2009).

In order to reveal the structural transformation processes of LAC and HPAEs, we make use of the product space, which is a network that formalizes the idea of relatedness between products traded in the global economy (Hidalgo et al., 2007; Hausmann et al., 2014). Moreover, we combine the product space with the Product

Gini Index (Hartmann et al., 2015) to reveal the relationship between a country's mix of products and its structural constraints on inequality reduction.

The Product Gini Index (PGI) is a measure recently introduced by Hartmann et al. (2015) that relates each product to its typical level of income inequality. Formally, the PGI is defined as the average level of income inequality of a product's exporters, weighted by the importance of each product in a country's export basket.

Finally, in this paper we introduce the *Xgini* as the average *PGI* of the products present on a country's portfolio. The *Xgini* aims to estimate the structural constrains to inequality imposed by a country's productive matrix. Formally, the *Xgini* of country $c$ is calculated as:

$$Xgini_c = \frac{1}{N_c} \sum_p M_{cp} s_{cp} PGI_p \qquad (1)$$

$$N_c = \sum_p M_{cp} s_{cp} \qquad (2)$$

where $s_{cp}$ is the share of product $p$ in the country's $c$ total export, $M_{cp}$ is 1 if product $p$ is produced by country $c$ with revealed comparative advantage and 0 otherwise, and $N_c$ is a normalizing factor to ensure that the *Xgini* is a weighted average of the PGI.

## 3 The gap in the productive capacities of LAC and HPAE

As previous work has shown, a country's mix of product can be seen as an expression of its institutions and the productive knowledge and know-how embedded in its society (Engerman and Sokoloff, 1997, Hidalgo, 2015, Hartmann et al., 2015). Here we compare the mix of products produced by LAC countries with that of HPAEs. The total export of all LAC countries is very similar to China's total export—$1.9 trillion vs. $2.2 trillion dollars in 2013, respectively. However, the differences in the productive capabilities between LAC and China become evident when looking at the types of products these economic regions export.

While a big part of China's total exports involves a gamut of manufactured goods, such as electronics, computer parts, or machinery, the percentage of manufactured goods in LAC countries' export portfolio is significantly lower. LAC economies export mainly raw materials and agricultural products, such as Crude Petroleum, Iron Ore, Copper, Coffee, and Soy Beans (Figure 1 A-B). The difference in the productive specialization and comparative advantages becomes even more pronounced when looking at the bilateral trade between these two parts of the world. LAC exports to China mainly raw materials, while China exports to LAC more sophisticated industrial products (Figure 1 C-D).

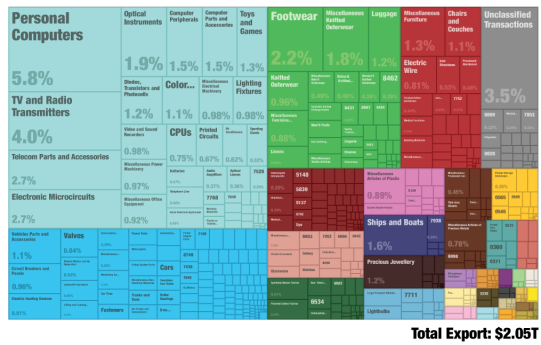
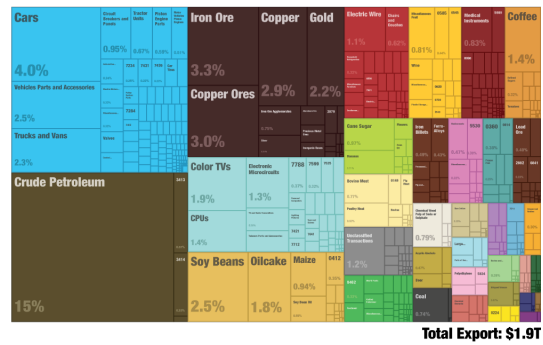
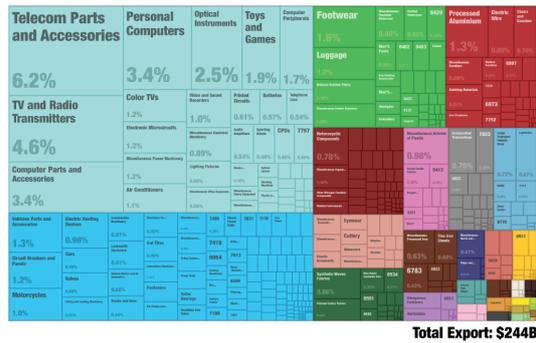
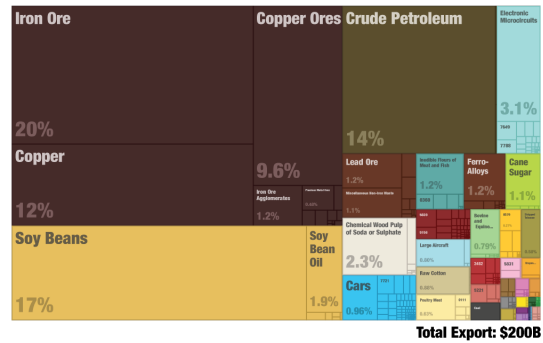

**Figure 1**. Export structure of **(A)** China to the world, **(B)** LAC to the world, **(C)** China to LAC and **(D)** LAC to China. Source: atlas.media.mit.edu

If we think of the productive sophistication of a country as an expression of the knowledge and knowhow embodied in its economy (C. Hidalgo 2015), then the trade pattern can be seen as an expression of the gap in knowledge and knowhow that exists between Latin American countries and China. The Economic Complexity Index (ECI) captures the differences in country's productive sophistication, taking both the diversity and the sophistication of a country's mix of products into account.

In the 2013 Economic Complexity ranking (Table 1), most LAC countries are significantly behind China (22th) and other Asian economies, such as South Korea (7th), Singapore (10th), Thailand (29th), or Malaysia (34th). The only outlier is Mexico, which ranks significantly higher than most LAC countries (23). But this is a fact that needs to be taken with reservations, since more than 70% of Mexico's exports are sent to the United States, suggesting that the apparent complexity of Mexico's economy is inflated due to its relationship with the U.S. Otherwise, we would expect a country with that level of productive sophistication to export to a larger number of destinations. Also in the case of Panama, the economic complexity index might be slightly overestimated as Panama has an important commercial free zone whose flows are usually mixed with the domestic ones (Ramos Martinez et al., 2015).

Table 1. Economic complexity ranking in 2013

**Top 5 countries**

| Rank | Country | ECI |
|---|---|---|
| 1 | Japan | 2.292 |
| 2 | Switzerland | 2.158 |
| 3 | Germany | 1.951 |
| 4 | Sweden | 1.827 |
| 5 | United Kingdom | 1.716 |

**HPAE**

| Rank | Country | ECI |
|---|---|---|
| 7 | South Korea | 1.699 |
| 10 | Singapore | 1.628 |
| 22 | China | 0.965 |
| 29 | Thailand | 0.758 |
| 34 | Malaysia | 0.693 |
| 49 | Philippines | 0.269 |

**LAC economies**

| Rank | Country | ECI |
|---|---|---|
| 23 | Mexico | 0.950 |
| 44 | Panama | 0.325 |
| 52 | Uruguay | 0.197 |
| 53 | Argentina | 0.187 |
| 54 | Colombia | 0.171 |
| 55 | Costa Rica | 0.162 |
| 56 | Brazil | 0.152 |
| 60 | El Salvador | -0.012 |
| 67 | Chile | -0.132 |
| 70 | Trinidad and Tobago | -0.188 |
| 74 | Jamaica | -0.331 |
| 76 | Guatemala | -0.377 |
| 79 | Paraguay | -0.418 |
| 80 | Dominican Republic | -0.421 |
| 85 | Peru | -0.553 |
| 87 | Honduras | -0.592 |
| 93 | Bolivia | -0.760 |
| 94 | Ecuador | -0.793 |
| 95 | Nicaragua | -0.810 |
| 99 | Venezuela | -0.908 |

**Bottom 5 countries**

| Rank | Country | ECI |
|---|---|---|
| 120 | Papua New Guinea | -1.670 |
| 121 | Mauritania | -1.702 |
| 122 | Libya | -1.712 |
| 123 | Turkmenistan | -1.753 |
| 124 | Guinea | -2.102 |

Source: atlas.media.mit.edu, own illustration

# 4 Structural transformation with equity

In this section, we compare the structural economic transformation of both regions to reveal the structural constraints and opportunities for inequality reduction. In the last decades, China and other HPAEs showed an increasing upward trend in their level of economic complexity (Figure 2-A). In LAC economic complexity has slightly increased until the debt crisis in the 1980s, and then remained or even declined in some cases (Figure 2-B). In the same time period, the Gini coefficients of countries such as Singapore, Thailand or Malaysia declined (Figure 2-C), while income inequality in LAC countries has increased since the 1980 (Figure 2-D). It must be noted that for China, reliable data is not readily available, though it seems that its level of income inequality has strongly increased (Xie and Zhou, 2014).

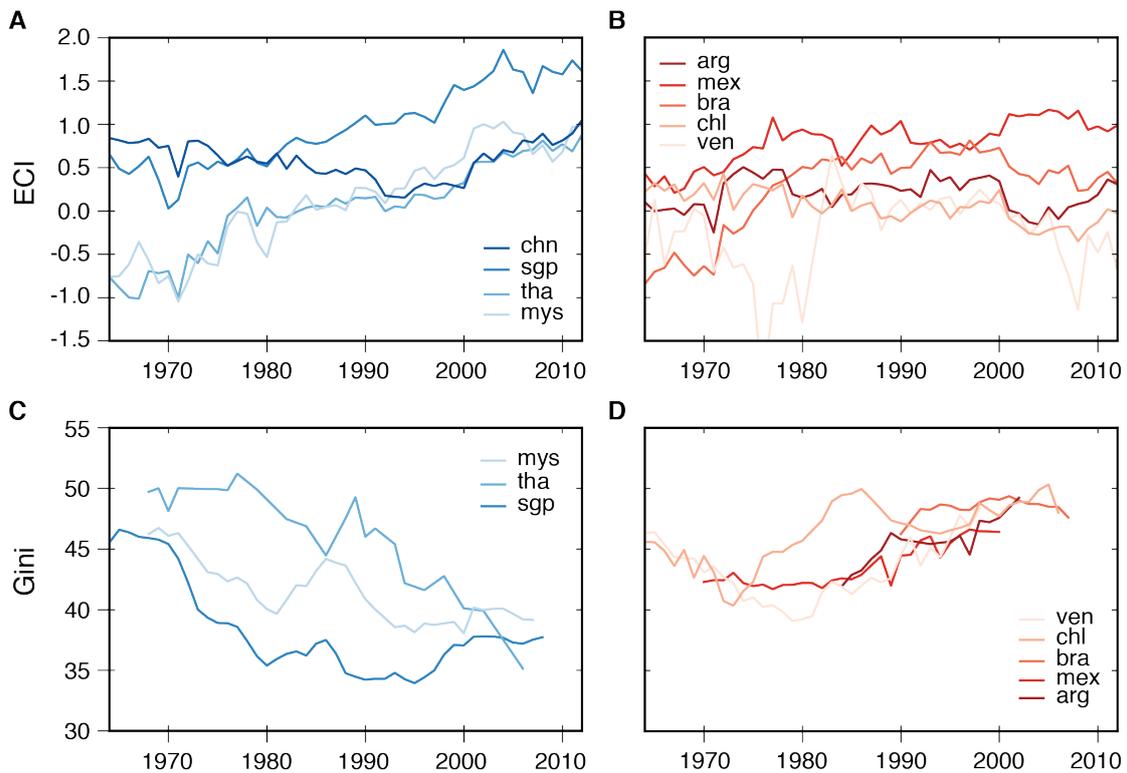

**Figure 2:** Evolution of Economic Complexity (panels A and B), and Income Inequality (panels C and D) in HPAE and in LAC countries.

However powerful the aggregate view provided by ECI and Gini might be, it cannot be used to illustrate the complex structural transformation processes and productive constraints. Hence, we overlay the product space of countries with the Product Gini Index (PGI) to get a better qualitative understanding about the kind of products countries produce, and the level of inequality associated to these products. This technique allows researchers, policy makers, and decision makers to gain structural insights into the development processes of their countries, and reveals the structural constraint on income inequality reduction imposed by their productive structure.

For example, in the 1970s both China and Brazil mainly exported products with a high PGI value, such as soybeans, tea, rice, linens, cocoa beans, wood, refined sugar, and crude petroleum (Figure 3). However, by the 2000s, China managed to diversify and also became competitive in a wide range of more sophisticated products like electronic and computer components and machinery, which are typically produced in countries with low levels of income inequality (—i.e. PGI values). Brazil, on the other hand, still depends on natural resources and the agricultural products it already produced in 1970s, such as coffee, soy beans, or refined sugar. Moreover, it can be observed that even though Brazil managed to expand some of its manufacturing industries, it also expanded its production in sectors associated with a high PGI values, such as the iron ore mining or tobacco production.

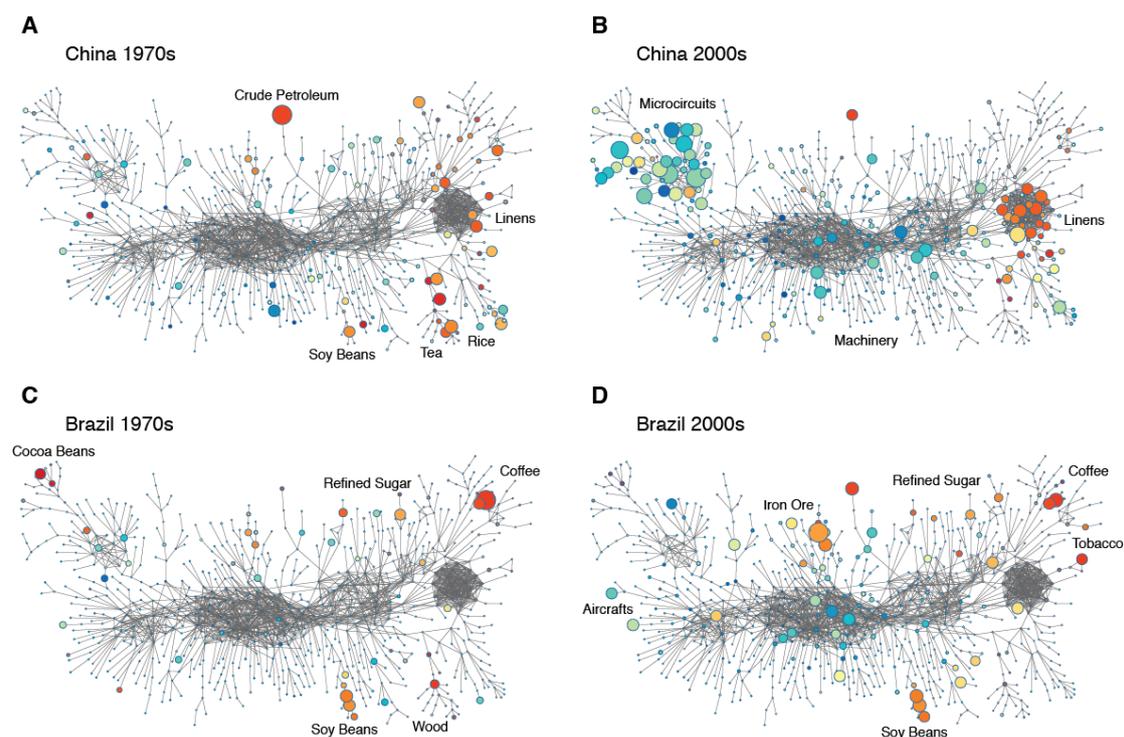

Figure 3. The structural transformation of China and Brazil

China and Brazil are not the only examples in which the different path of the productive transformation of HPAE and LAC countries becomes evident. Between the 1970s and the 2000s, South Korea managed to transform almost their entire export portfolio into more sophisticated products, such as cars, hydrocarbons, and polyethylene. It managed to move away from the products with high PGI that it produced in the 1970s into products with lower PGI values. During the same time period, Peru barely diversified (Figure 4) and is still dependent on products with high PGI values such as copper, iron ore, and fish.

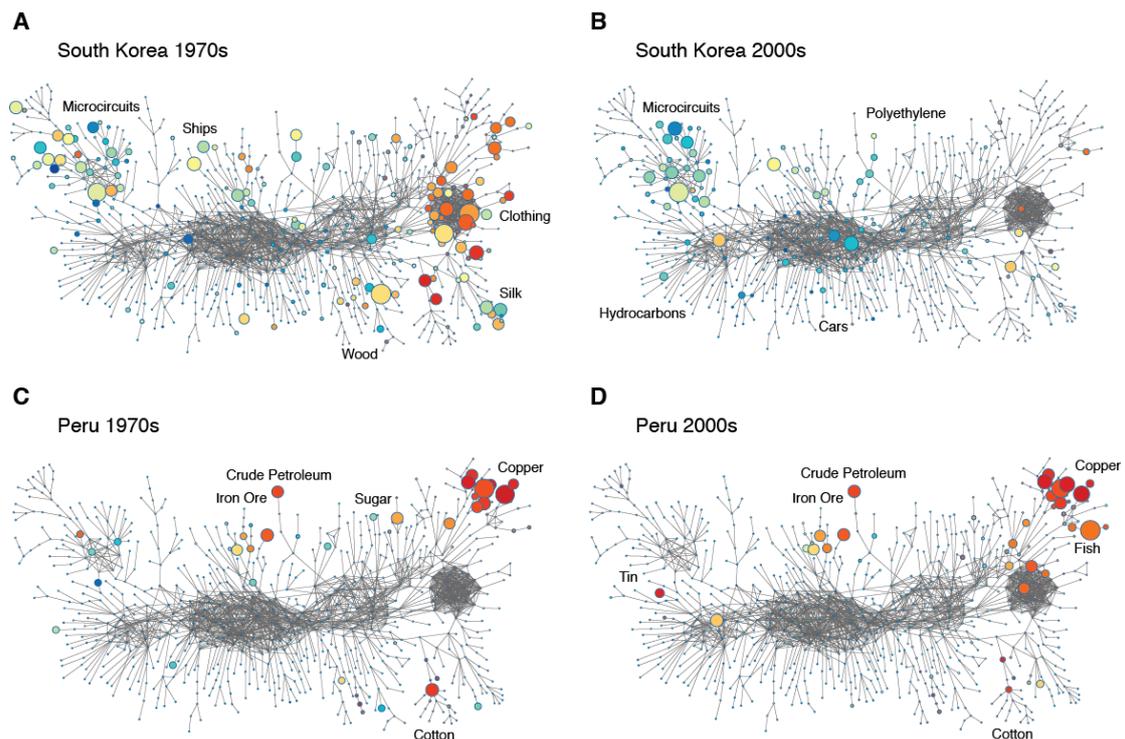

Figure 4. The structural transformation of South Korea and Peru

We can also use the information about the comparative level of income inequality related to different types of products (i.e. PGIs) to estimate the structural constraints on inequality imposed by a country's mix of products. Based on the PGIs, we can calculate the Xgini, which estimates the level of inequality expected from a country's productive structure. Of course, deviations of the Gini from the Xgini might be a consequence of many other processes which are not related with a country's productive structure, such as social policies, tax regimes, etc. However, different types of products and productive structures also tend to be strongly associated with different types of institutions, human capital and the level of income inequality, as a vast amount of literature on economic development has shown (Furtado, 1959, Innis, 1970, Engerman and Sokoloff, 1997, Collier, 2007, Hartmann, 2014, Hartmann et al., 2015). A more diverse and sophisticated productive structure creates more opportunities for labor mobility and bargaining power of the workers, favors a better distribution of economic and political power, and is associated with more inclusive institutions (ibid.).

Figure 5 shows the evolution of the average Xgini for LAC and Asian economies. The HPAE's average Xgini has significantly declined since the mid 1980s, implying that these countries have diversified into products related with lower levels of inequality. As a result of these transformations, HPAEs have been able to generate a large amount of job opportunities in new industries. While China and other HPAEs created the potential for a more complex and inclusive economy, most LAC countries have not yet created these opportunities at the same scale. Instead, the Xgini of most LAC economies has remained the same—with the exception of Mexico. In other words, LAC countries continue to export products associated with high levels of inequality and low levels of economic complexity.

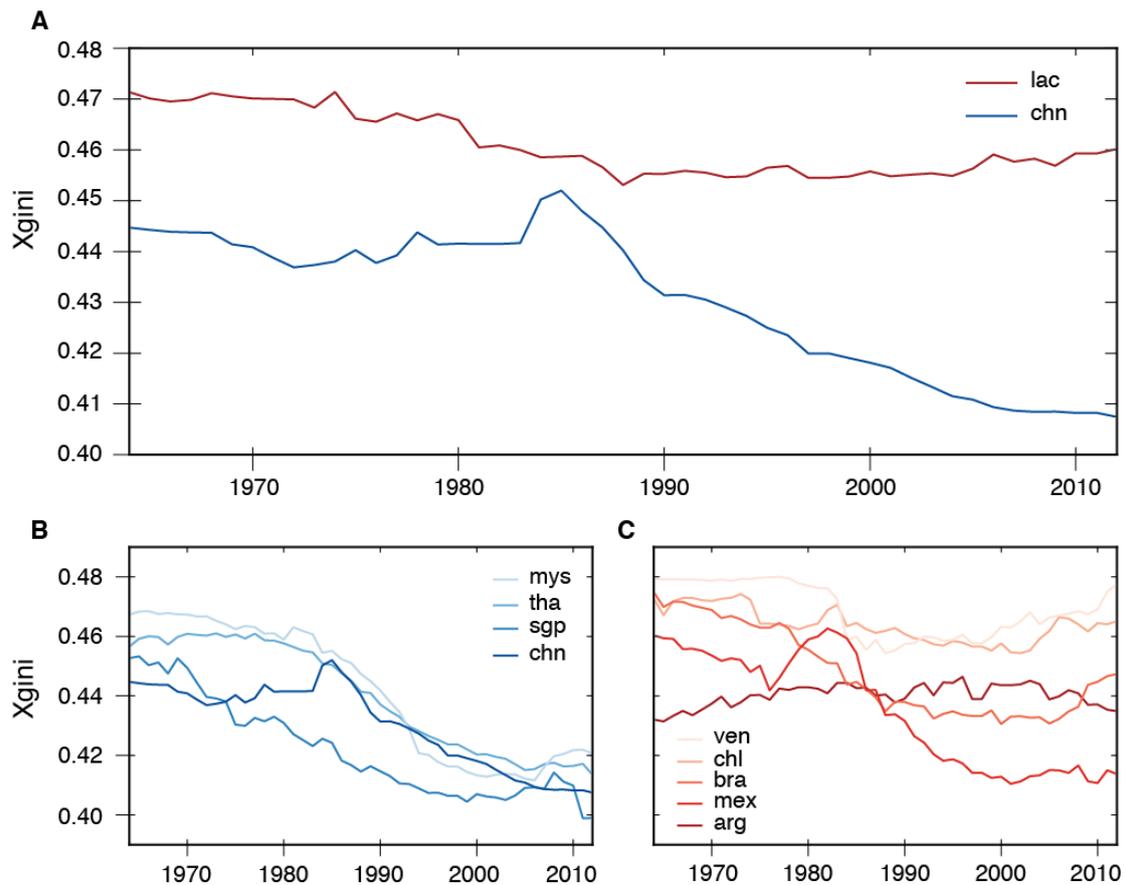

Figure 5: The evolution of Xginis in HPAE and LAC countries between 1962-2013. (A) Comparison of the Xgini of China with the average Xgini of LAC countries excluding Mexico. (B) Xginis of HPAE (C) Xgini of LAC countries

## 5 Economic policies in HPAE and LAC

While the economic policies of LAC countries have oscillated been advocates of strong state intervention or complete liberalization and deregulation, HPAE are good examples of a more successful middle ground, introducing both market forces and strong government investment in human capital and innovation. The double emphasis on both social and industrial policies has helped the HPAE to promote technological upgrading and increase their economic complexity.

The profound structural transformation and increase in economic complexity of HPAE has led to profound changes in their distribution of income. Through the concerted mix of social and economic policies, smaller to middle sized HPAEs, like Singapore, South Korea, Malaysia or Thailand, have been able to spread the benefits of the increase in economic complexity and effectively decreased their income inequality (Stiglitz 1996, Wade, 1990). In the case of China the economic reforms, and associated increase in economic complexity, have helped hundreds of millions of Chinese out of extreme poverty. Due to several factors, though, such as the spatial concentration of the economic activity and the urban-rural divide, it has also seen a strong increase in its level of income inequality (Xie and Zhou 2014). Nonetheless, the increase in economic complexity and ongoing technological catch-up is also creating the opportunities for further employment generation and the potential to further spread social welfare. Moreover, the rise of new industrial sectors in China has generated the need for institutional changes to address societal problems like urbanization, ecological problems or migrant workers. If China manages to address these challenges by spreading technological progress and implementing inclusive institutions across its vast country, then—in line with the arguments of the Kuznets' curve (Kuznets 1955)—this process can lead to a decrease in income inequality. Thus, China is certainly still facing many difficult challenges, but its increase in economic complexity has also opened up opportunities for further inclusive growth, reducing both absolute and relative income inequality.

The same economic opportunities for inequality reduction are not yet present in most of LAC economies, since they are still constrained by productive structures centered on natural resources. While HPAE are rapidly catching-up and forging ahead in several technological and productive areas, most LAC countries are still dependent on a much smaller set of productive activities that provide a narrower set of new job opportunities. As shown by several works in institutional and development economics, the historical pattern of economic specialization in resource-exploiting activities in LAC countries has undermined its capabilities for inclusive growth and led to high levels of income inequality, and often to exploitative institutions (Engerman and Sokoloff 1997; Acemoglu and Robinson 2012; Hartmann et al. 2015). Moreover, in contrast to other historically resource-intensive economies like Norway, Canada or Australia, LAC's economies have been less successful in establishing more inclusive institutions and generating related complex industries like harvesting machinery, advanced metal products, newsprint or paper making machinery.

In recent years, several LAC countries have been able to reduce poverty and inequality by raising social expenditures and implementing several social policy programs—like the conditional cash-transfer-program "Bolsa Familia" in Brazil. Measures like these have also increased the average years of schooling and life expectancy. However, despite comparatively high levels of years of schooling, life expectancy, and active social interest groups, most of the LAC countries have not managed to significantly change their productive matrix. In consequence, the increase in education and human development has not been matched by economic opportunities for the labor force to work in more knowledge based and complex industries.

If LAC countries are ever to significantly reduce their levels of income inequality, the main remaining challenge is to match the social policies with industrial policies that facilitate higher levels of economic complexity based on opportunity based entrepreneurship and innovation systems. In order to create prolific innovation systems, social and industrial policies need to complement each other in a congruent way. Therefore, it is important to overcome the still persistent state versus market battle in the Latin American policy debate. Effective innovation and industrial policies involve both market forces and state intervention to raise human capital, address both market and government failures, establish innovative industrial clusters and promote interactive learning between all the agents involved in the economy (Lundvall 2010; Giuliani, Pietrobelli, and Rabellotti 2005, Hartmann, 2014). In order to achieve the difficult task to overcome the structural constraints and increase the economic complexity of Latin America, companies, government agencies, academia, and the civil society need to work together and learn from each other in order to spread knowledge, introduce innovations and increase economic complexity (Lundvall et al., 2011; Hartmann, 2014). In order to create and establish new industries, the economic agents need to be enabled to figure out which products work best in their regions or countries in a self-discovery process (Hausmann and Rodrik, 2003; Hartmann, 2014). It is important to note, though, that a single country cannot produce all inputs in a competitive way in the modern globalized economy. Thus when figuring out economic opportunities economic agents also need to deliberately search for and access international markets and knowledge sources, for example through, commuting entrepreneurs and outwards oriented development strategies (Saxenian 2007; Pyka, Kustepeli, and Hartmann 2016).

# 6  Concluding remarks

This paper adds to the mounting evidence that the mix of products a country exports is important for its economic development in terms of GDP (Rodrik 2006; Hausmann, Hwang, and Rodrik 2006; Hidalgo and Hausmann 2009; Hausmann et al. 2014; C. Hidalgo 2015) and income inequality (Hartmann et al., 2015). Here, we have used methods from economic complexity research to reveal the gap in productive capabilities and opportunities for inequality reduction of LAC and HPAE. The results show that HPAE have been capable of increasing their level of economic complexity and thus overcoming structural constraints on income inequality—as expressed in the decline of their Xgini—. Conversely LAC are still strongly constrained by its natural-resource centric productive structures and its Xgini has remained on a high level.

Despite the recent positive impact of successful social policy programs in Latin America, without also simultaneously raising the level of economic complexity, social policies may lack the strength required to modify a country's level of income inequality beyond what is expected due to its productive structure. In consequence, to have sustained economic growth and reduction of income inequality in LAC, prolific industrial policies that complement social policies are necessary (Amsden 2010; Hartmann 2014; Hartmann et al. 2015). This also implies the need to overcome polarizing state versus market policy debates and implement a smart combination of both market forces and state subsidies that are able to promote prolific innovation systems and raise economic complexity in Latin America and the Caribbean.

**Acknowledgements:** All authors acknowledge the support from the MIT Media Lab consortia. Dominik Hartmann acknowledges support from the Marie Curie International Outgoing Fellowship No. 328828 within the 7[th] European Community Framework Programme. Miguel R. Guevara and César A. Hidalgo acknowledge support from MIT-Chile MISTI fund.